# Triggering InAs/GaAs Quantum Dot Nucleation and growth rate determination by in-situ modulation of surface energy

Peter Spencer*, Chong Chen, Wladislaw Michailow, Harvey Beere, David Ritchie

Cavendish Laboratory, Cambridge University, JJ Thomson Avenue, Cambridge CB3 0HE, UK

*contact e-mail address: peter@quantized.art

## Abstract

Epitaxial InAs/GaAs Quantum Dots (QDs) are widely used as highly efficient and pure sources of single photons and entangled photon-pairs [1, 2, 3], however reliable wafer-scale growth techniques have proved elusive. Growth of two-dimensional Quantum Well (QW) thin-films can be achieved with atomic precision down to below the de Broglie wavelength of electrons in the material, exposing the quantum "particle-in-a-box" energy vs. thickness-squared relationship [4]. However, difficulties in controlling the exact moment of nanostructure nucleation obscure this behaviour in epitaxial QD material, preventing a clear understanding of their growth. In this work we demonstrate that QD nucleation can be induced by directly modulating the crystal's surface energy without additional materials or equipment. This gains us quantitative measure of the QD growth rate and enables predictive design of QD growth processes. We believe this technique will be crucial to the realisation of uniform arrays of QDs required for scalable quantum devices at the single-photon level.

## Introduction

The key to a good and even surface coating, be it paint or semiconductor, is that the deposited layer has good wetting characteristics (top, Fig. 1). Thermodynamically, wetting occurs when depositing a layer on a substrate without increasing the surface energy:

$$E_\text{epilayer} \leq E_\text{substrate}$$

(Equation 1)

When the epilayer's surface energy $E_\text{epilayer}$ exceeds the substrate surface energy, $E_\text{substrate}$, deposited material adheres to itself more than the substrate, forming an irregular, possibly detached layer (left, second row in Fig. 1) [5]. To achieve a successful paint finish, for example, the condition in Eq. 1 above must be maintained until the paint dries to avoid a phenomenon called dewetting, which would ruin the surface finish (right-to-left, second row, Fig. 1); however, dewetting finds a use in creating crystalline nanoparticles [6]. Stranski and Krastanow discovered that when one crystalline material is deposited on a dissimilar crystal substrate, the condition of Eq. (1) can be satisfied for the first layer (the "wetting layer", WL), but not for the following layer [5, 7]. Under certain conditions however, the balance of Eq. (1) can still be satisfied for multiple deposited layers but with a gap between the stable WL and the subsequently stable layers. This turns out to allow the self-assembly of nanostructures.

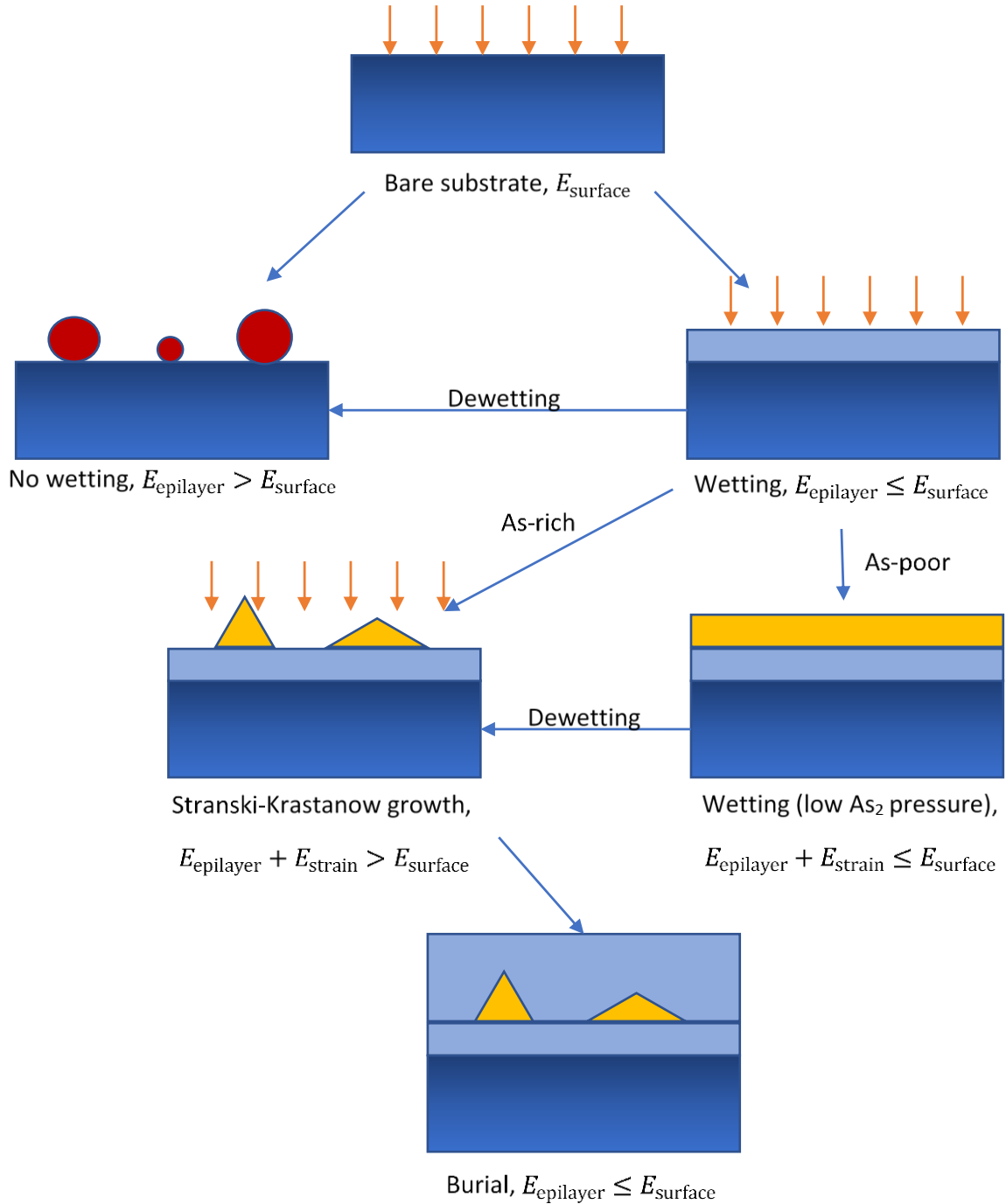

*Figure 1: The stages of InAs/GaAs QD epitaxial growth, starting with bare GaAs substrate (top), through the first layer deposition (second row), subsequent layer deposition (third row), and an optically-active buried QD layer (bottom). Wetting is required to deposit a planar layer but dewetting can also occur with changes in conditions and damage the layer (second row). Stranski-Krastanow growth occurs when the first layer(s) wet completely but subsequent layers do not (diagonal to third row), leading to nucleation of QDs; however, modulation of the surface energy can also induce QD nucleation by dewetting: giving a well-defined and controllable start to QD growth.*

When InAs is deposited on a GaAs substrate, elastic strain energy builds up because of the 7% strain mismatch between their bulk lattices. Eq. (1) is initially satisfied, resulting in a smooth WL; then as further InAs is deposited the increasing strain causes $E_{\text{epilayer}}$ to rise until it exceeds $E_{\text{substrate}}$, inducing dewetting at a critical coverage of InAs, $\theta_{\text{crit}}$ (third-row, Fig. 1). The dewetting results in nucleation of nanometre sized islands without dislocating the crystal lattice. Subsequent burial of these islands (bottom-row, Fig. 1) provides three-dimensional confinement of electrons and holes, meaning they behave as "artificial atoms" with discrete energy levels whose properties depend on their size [8]. Isolating the optical transitions from individual QDs allows the creation of single- and entangled-pair photon sources: This requires precise control of the size and number of nanocrystals, which has proved challenging to achieve with methods that can scale to production volumes. Here we demonstrate the nucleation process can be precisely and independently controlled with standard Molecular Beam Epitaxy (MBE) reactors and confirm that QDs grow by linearly increasing in volume in direct proportion to time; the result is a reliable and scalable QD growth process for Quantum Communication applications.

## Wetting Layer surface energy

Microscopic studies of the WL surface during InAs deposition on GaAs have revealed a complex mixture of InAs, GaAs, and InGaAs alloyed domains [9]. Crystal surface structure can differ significantly from the bulk lattice because the periodicity of the crystal potential is interrupted, in this case the surface structure is "reconstructed". Multiple surface reconstruction unit cell configurations can co-exist, and unlike the bulk unit cell, these can be non-stoichiometric with different number densities of group-III (Ga: $N_{\text{Ga}}$, In: $N_{\text{In}}$) and group-V (As: $N_{\text{As}}$) atoms, giving rise to a complex WL surface. Surface reconstruction is monitored in-situ by Reflection High-Energy Electron Diffraction (RHEED) measurements. During InGaAs deposition by MBE, a RHEED pattern with $(n \times 3)$ symmetry is typically observed due to As-rich surface domains of InGaAs which have been correlated with QD nucleation [10].

Non-stoichiometry of As-rich surface reconstruction domains is important here because, from the definition of chemical potential, an imbalance of group-III and group-V atoms makes the surface sensitive to changes in the As chemical potential $\mu_{\text{As}}$:

$$\Delta E_{\text{surface}} = (N_{\text{Ga}} + N_{\text{In}} - N_{\text{As}})\Delta\mu_{\text{As}}$$

(Equation 2)

$\mu_{\text{As}}$ is controlled during MBE growth by varying the substrate temperature $T_{\text{sub}}$ and the flux of As$_2$ molecules impinging on the surface (the Beam Equivalent Pressure, BEP). $T_{\text{sub}}$ determines the effusion rate of As$_2$ from the surface, which is necessary for a well-defined value of $\mu_{\text{As}}$ to exist. By lowering $T_{\text{sub}}$ or raising the As$_2$ BEP it should be possible to reduce $E_{\text{surface}}$ and tip the balance of Eq. (1) for the As-rich $(n \times 3)$ surface, inducing dewetting and nucleating QDs (third row, Fig. 1). Mirin *et al.* observed the dewetting transition upon lowering $T_{\text{sub}}$, confirming that this is possible [11]. Therefore, it should be possible to separate the deposition of InAs from QD nucleation by controlling the As$_2$ BEP.

Variable control of As$_2$ BEP is straightforward in modern MBE systems equipped with a valved-cracker As$_2$ source. To trigger the dewetting transition by varying the As$_2$ BEP, varying amounts of InAs were pre-deposited under As-poor conditions prior to re-opening the As$_2$ valve to its normal position, while monitoring the RHEED signal (see Appendix). Figure 2(a) compares the RHEED data for continuous and "triggered" QD nucleation: increasing the As$_2$ BEP causes a sudden formation of QDs on the surface, in contrast to the gradual onset of nucleation under conventional continuous InAs deposition. Figure 2(b) shows the repeatability of the process under identical conditions. Figure 2(c) shows that when varying the pre-deposited coverage of InAs $\theta$ above $\theta_{\text{crit}}$, QD nucleation occurs at the same moment after opening the As$_2$ valve. This shows control of the precise timing of QD nucleation, independent of InAs coverage $\theta$.

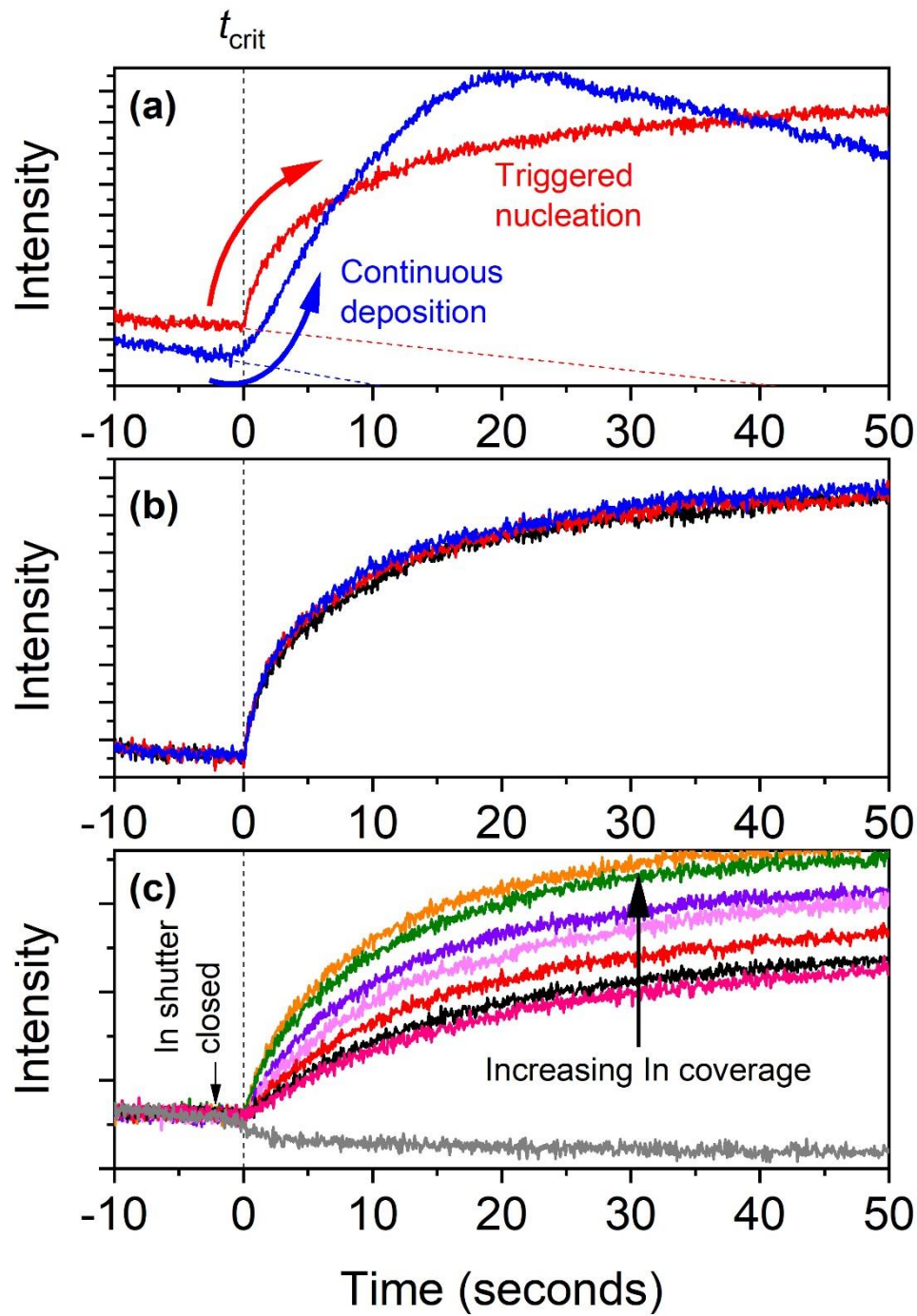

*Figure 2 (a): A comparison of the RHEED signal during continuous InAs deposition (blue line) and triggered QD nucleation (red line) by opening the As$_2$ source valve. There is a clear difference between the gradual vs. sudden onset of nucleation at 0 seconds, respectively. (b) The RHEED signals from three repeats under identical conditions on the same wafer show excellent repeatability. (c) The results of varying the pre-deposited InAs coverage $\theta$: QD nucleation is observed at the same time for each RHEED measurement, where the coverage was changed by varying the InAs pre-deposition time. The lower trace is a reference where $\theta < \theta_{crit}$ InAs was pre-deposited and no QD nucleation occurs on opening the As$_2$ source valve.*

# Growth rate

For thin-film QW layers the start, rate, and end of deposition are well-controlled; the quantum confinement effect can then be verified by using optical measurements to observe the energy eigenstates' dependence on layer thickness $L_z$ [4]. In contrast, the rate of self-assembly of each QD nanostructure depends on the local conditions, proceeding until some limiting equilibrium size, or the plastic deformation limit is reached. The progressive onset to QD nucleation with conventional strain-driven nucleation of QDs has no well-defined start. As a result, until now a layer of QDs develops as a random population and a statistical approach is required.

By modulating the As$_2$ BEP to trigger the nucleation point using the method outlined above, the start of growth is now well-controlled. We can now investigate the growth-rate: To do this a well-defined end to QD growth is required, and the only option is to bury the nanostructure by depositing GaAs and measure the QDs optically. Burial, or capping, limits further development of the QD and while this process also causes an erosion of the underlying nanocrystal, it is required for most device designs that incorporate InAs QDs: optical measurements (such as PL) will therefore provide direct correlation between growth and the final QD properties.

A series of samples was prepared where a layer of InAs was pre-deposited on a GaAs substrate under As-poor conditions, $P_{As} = 1.1 \times 10^{-7}$ Torr. After closure of the In shutter, the As$_2$ source valve was commanded to open and then after a pre-set time-delay $t$, the Ga shutter was opened to begin the capping process (see Appendix). Precise control of $t$ is essential. In each series, three samples were grown with time delays of $t = 3$ s, 4 s, and 5 s. For the first series, the valve on the arsenic cracking source was actuated slowly at a velocity of $R_{slow} = 3.33\%$ s$^{-1}$, while for the second series the valve was actuated faster at $R_{fast} = 40\%$ s$^{-1}$, as shown at the bottom of Figure 3. The sample series were denoted by "S$t$" and "F$t$" for the "slow" and "fast" series respectively, where $t$ is as defined above.

Initial photoluminescence (PL) measurements of the ensemble properties of the QDs were done in a microscope with a low spatial resolution. Small pieces of the wafer were cleaved, mounted in a Janis ST500 cryostat and cooled to 5.2 K under vacuum. The 632.8 nm line of a Helium-Neon laser was used to excite the sample by focusing through a 5X Mitutoyo long working distance objective that also collected the resultant PL. Detection of the PL was achieved by dispersing the light through a 150 lines per mm grating in an 0.5 metre monochromator onto an InGaAs array detector. The ensemble PL spectra are shown in Figure 3(a), acceptor emission is evident at 830 nm and WL emission peaks at 879 nm. Samples from the slow-valve series exhibit greater WL emission, suggesting either a lower density of QDs, or fewer non-radiative centres. QD emission is visible from 879 nm (verified by micro-PL) to 1000 nm. The QD ensemble has an asymmetric distribution that appears to be a similar shape in all samples, growing broader with increasing $t$.

A QD's height $L_z \lesssim 5$ nm is smaller than its width ($\gtrsim 15$ nm) [12], thus the confinement energy arises mostly due to confinement in growth direction. This allows the use of a one-dimensional finite QW model to estimate QD volume (see supplemental information), since the pyramidal shape of an InAs/GaAs QD is known to be relatively constant during growth [13]. Thus, by considering that the volume $V = \alpha L_z^3$ (where $\alpha$ as a factor accounting for the QD aspect ratio), together with the solution to the one-dimensional Schrödinger equation for $L_z$, allows the transformation of the wavelength axis into an "effective QD volume", $V$.

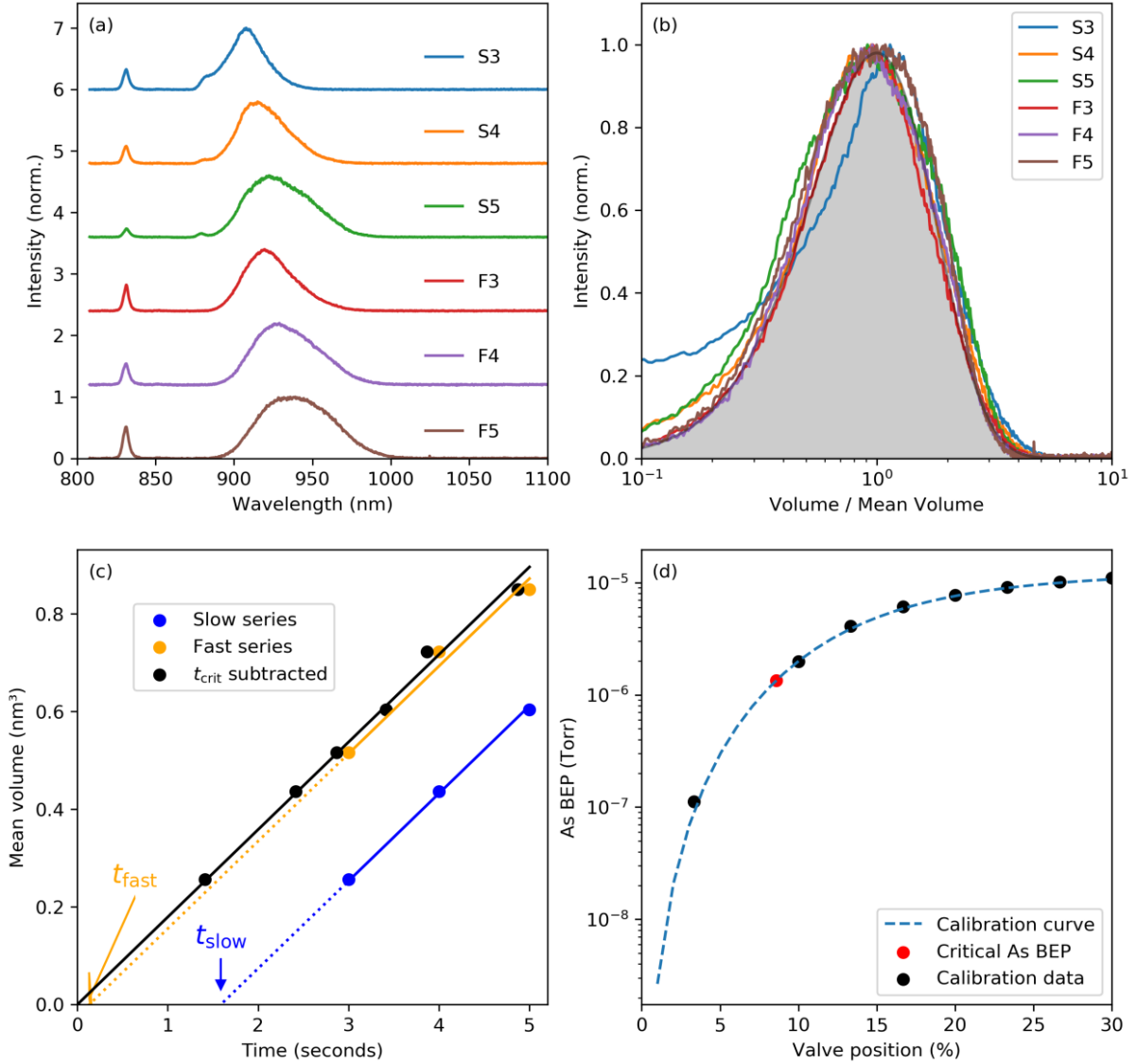

*Figure 3: (a) The raw PL data from each sample. Carbon acceptor luminescence is observed around 830 nm, WL emission is seen peaking at 879 nm, co-incident with the onset of QD ensemble emission; the QD emission redshifts and broadens with increasing growth time t. (b) After transforming the wavelength axis into volume and normalising for the mean, the PL from each sample has a consistent gamma-distributed shape. The shaded area is the theoretical distribution with shape factor $\beta = 3.61$, expected for randomly-positioned nanostructures on a surface. (c) The mean effective volume for each sample, plotted against time. The blue and yellow points are for the slow and fast samples, respectively. Fitting a straight-line to each data set infers the nucleation time $t_{crit}$ for the slow and fast series ($t_{slow}$ and $t_{fast}$ respectively); once $t_{crit}$ is subtracted (black points), the QD growth is linear in volume as a function of time after nucleation. (d) The $As_2$ BEP calibration data, showing the critical arsenic pressure for QD nucleation.*

Figure 3(b) shows the results of applying our model's transformation into $V$, normalised by the mean volume $\bar{V}$, showing the distribution of QD volumes is similar for each sample; a stable distribution shape indicates that the growth-rate of each QD is determined by its size, i.e. growth is scale-invariant in the early stages after nucleation. The QD populations all follow a Gamma distribution and the shaded area in Figure 3(b) is the shape expected for growth of randomly positioned particles governed by diffusion on a two-dimensional surface (see supplementary information) [14, 15].

$\bar{V}$ versus $t$ is plotted in Figure 3(c) as open symbols. Fit lines are plotted for the fast and slow sample series with a common growth-rate of $(0.179 \pm 0.004)$ nm$^3$s$^{-1}$ (with $\alpha = 1$), showing excellent agreement with the data; the actual growth-rate of the QDs will be much larger due to the higher aspect ratio of a real QD leading to $\alpha \gg 1$ and the erosion of QD height during capping, as well as In-Ga intermixing within the QD; for example a QD with 20 nm base length and 5 nm height with pure InAs composition may have $\alpha \geq 80$

We expect QD nucleation when the balance of $E_{\text{epilayer}}$ and $E_{\text{substrate}}$ reverses, and that this point depends on the As$_2$ BEP, $P_{As} = P_{\text{crit}}$, and $T_{\text{sub}}$ which was held constant. Comparing $x$-intercept timings with valve opening rates and the As$_2$ BEP calibration curve Figure 3(d), we obtain a simultaneous equation whose solution determines that QD nucleation occurs when $P_{\text{crit}} = 1.34 \times 10^{-6}$ Torr at $T_{\text{sub}} = 475$ °C.

The density of QDs in each layer was measured to average 1.2×10$^{10}$ cm$^{-2}$ from Atomic Force Microscopy (AFM) of uncapped layers grown by the same method. Assuming this density is comparable to the buried layers, this is enough information to predict the QD growth time required to obtain a useable optical density of QDs at a chosen target wavelength. Figure 4(a) shows predicted lines for a constant optical density of QDs against $t$, with points plotted to show the wavelength range where PL from individual QDs were easily resolved using a 50X Mitutoyo objective (numerical aperture 0.55), as shown in Figure 4(b). Full details of this calculation and supporting PL data are in the supplementary information.

To verify isolated QD emission, a study of excitation power dependence was carried out and a representative result is shown in Figure 5(a). To prove that the emission is coming from solitary QDs, the exciton and biexciton emissions were identified as shown in Figure 5(b).

# Conclusion

We have demonstrated controlled nucleation of self-assembled InAs/GaAs QDs, independently of InAs coverage above the critical thickness $\theta > \theta_{\text{crit}}$, using standard MBE growth. This allowed the linear growth rate of volume with time to be observed from PL data, showing that wafers of QD material with isolated QD emission at precise wavelengths can be designed and grown with unprecedented repeatability. In our early tests, isolated QD emission of wavelengths greater than 1000 nm were observed.

The predictability of the growth brings benefits both in designing optimised growth procedures, and in rapid characterisation feedback of wafer material: a simple PL measurement or wafer-mapper can observe the ensemble distribution of many QDs to fully parameterise the individually-resolvable QD emission. This makes production of single-photon source wafers a straightforward proposition.

The simple growth technique as presented is ideally suited to the reliable wafer-scale production of "single-QD" material with randomly-positioned QDs because it does not require a rotation-stop, but it is limited by the large overall density of QDs and the lack of position control. Optimisation of pre-deposited coverage of InAs is likely to gain control over density but we expect that site-control of QDs is the ultimate application: This technique of triggered nucleation will allow full control of nanostructure size when combined with pre-defined nucleation centres and well-controlled local environments, minimising inhomogeneous broadening and offering the possibility of ordered arrays of identical QDs.

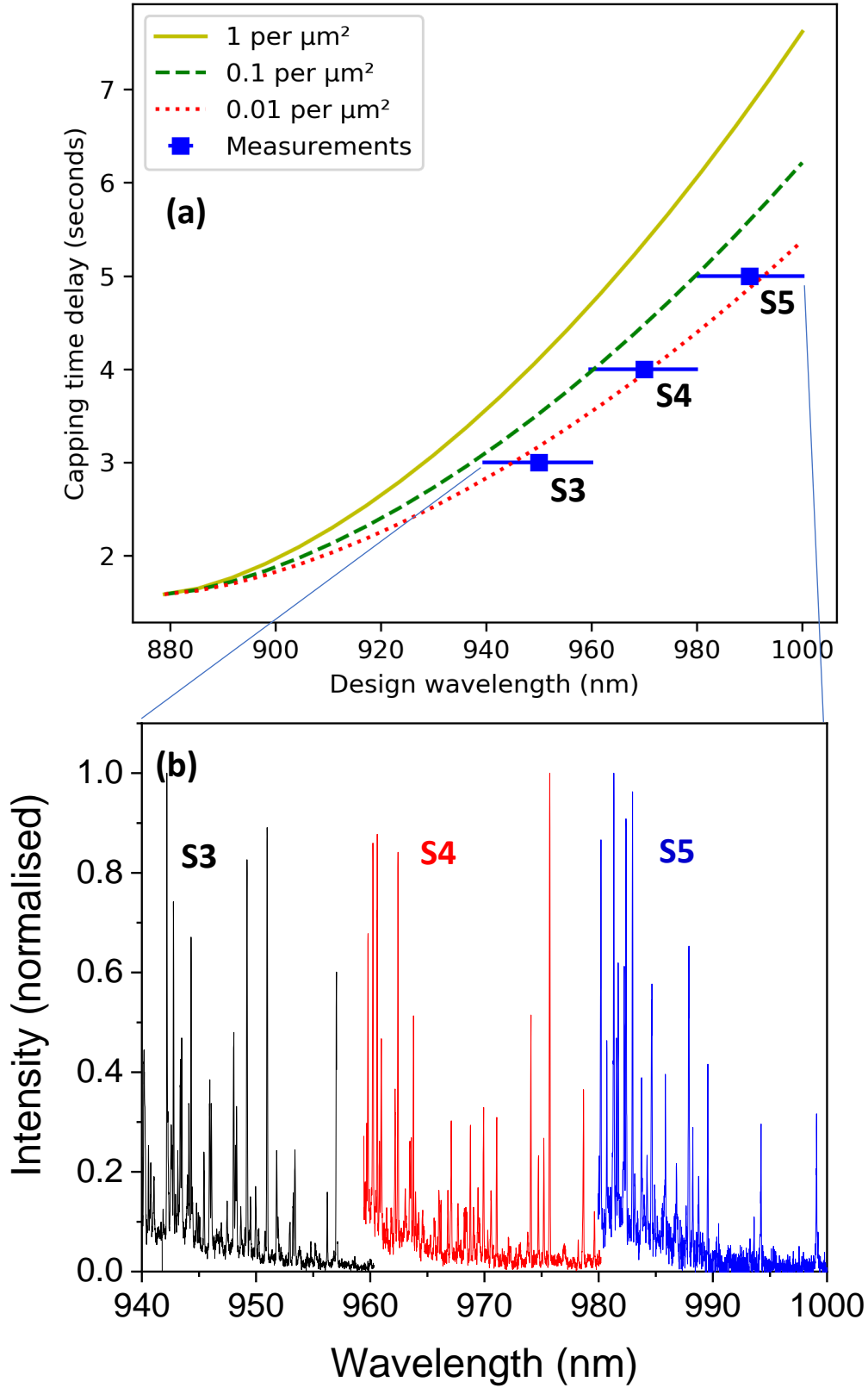

Figure 4: (a) Lines predicting QD growth time for a given wavelength and density, points are measured. (b) Representative spectra for each of the slow series samples, taken in a microscope with estimated 1/e collection area of 13 µm$^2$ and showing resolved QD emission as predicted.

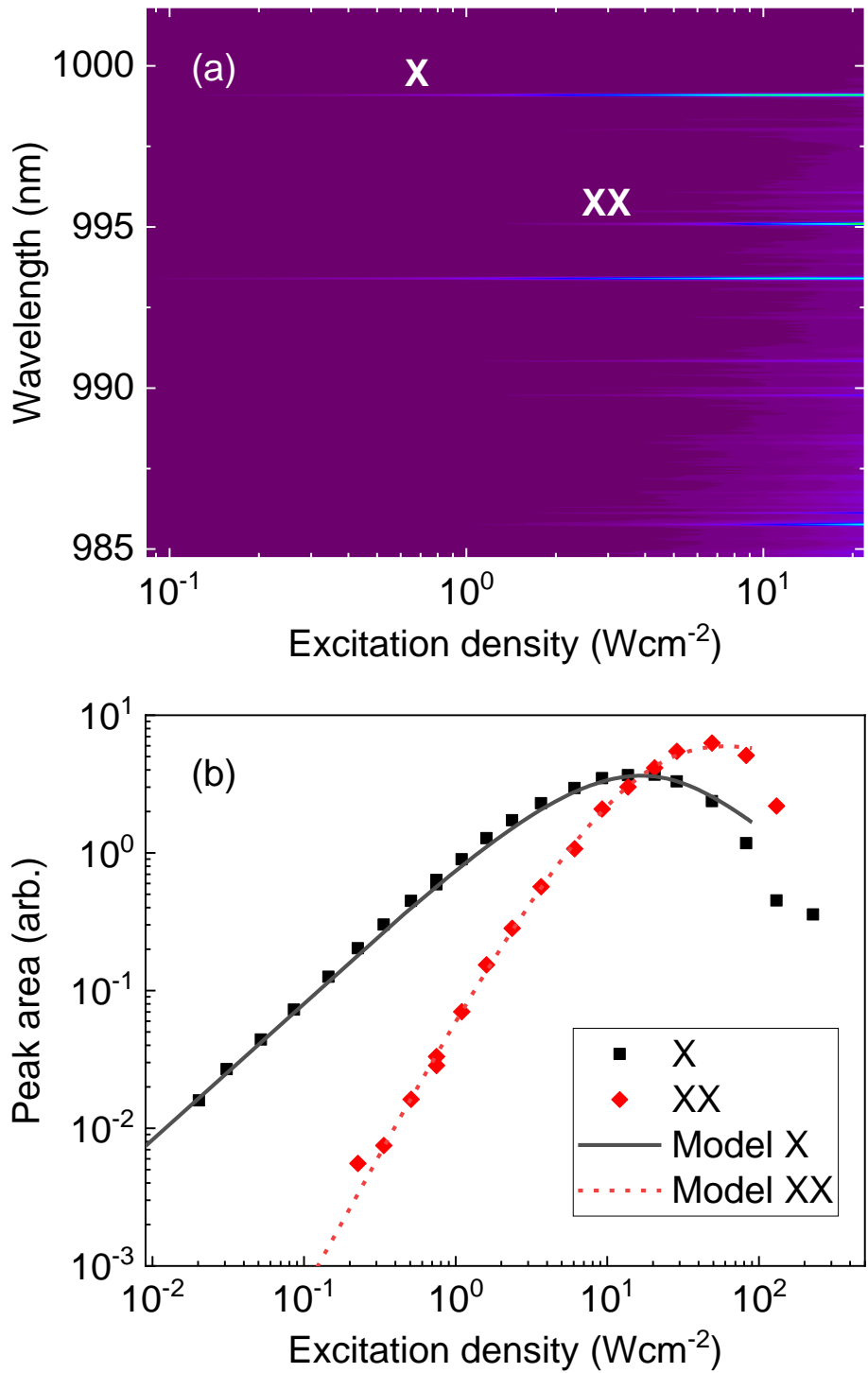

*Figure 5: (a) PL results on isolated QD emission from sample S5 taken with a 50X Mitutoyo objective, the colour scale indicates PL intensity. Emission from a neutral exciton X at 999.1 nm, and the corresponding biexciton emission XX at 995.1 nm are labelled. (b) The power dependence of the identified X and XX lines, with the solid and dotted lines showing the expected behaviour for the exciton and biexciton predicted by a simple model (see supplementary material). Below 3 Wcm$^{-2}$ the exciton and biexciton have linear and parabolic dependences on excitation density, respectively.*

# Acknowledgements

Peter Spencer acknowledges funding by the E.P.S.R.C. Quantum Communications Hub (grant number EP/M013472/1) and is grateful for discussions with and the support of Dr Daniel Farrell. Chong Chen acknowledges funding from the E.P.S.R.C through grant number EPSRC EP/K004077/1. Wladislaw Michailow thanks the George and Lillian Schiff Foundation of the University of Cambridge for financial support and is grateful for the Honorary Vice-Chancellor's Award of the Cambridge Trust.

# Appendix: Epitaxial growth

## Initial sample preparation

MBE growth was performed on (001) GaAs semi-insulating wafers that were first heated to 620 °C for ten minutes to remove the oxide layer, before being cooled to 580 °C for the deposition of a GaAs buffer layer of at least 200 nm thickness at a rate of 1 μmhr$^{-1}$, rotating the sample at 30 RPM and with an As$_2$ BEP of $P_{As} = 6 \times 10^{-6}$ Torr. QD growth then proceeded according to the schematic timeline illustrated in Figure 6: After the Ga shutter closed, the surface was then annealed under an As$_2$ flux at 580 °C for a further 10 minutes before cooling to 475 °C for InAs pre-deposition. During the cooldown, the As$_2$ source valve was closed to 3.3% to reduce the As$_2$ BEP to $P_{As} = 1.1 \times 10^{-7}$ Torr, suppressing QD formation during InAs pre-deposition. Once the substrate temperature had settled, the In shutter was opened to continuously deposit a selected coverage of InAs, $\theta$, at a rate of 0.023 MLs$^{-1}$ (Fig. 6b). Once $\theta$ InAs was deposited, the In shutter was closed and the As$_2$ source valve opened (Fig. 6c,d). At this point, either RHEED measurements were performed or a PL/AFM sample prepared as discussed below.

## RHEED measurements

RHEED measurements were performed using a 15 keV Staib electron gun, reflecting from the substrate onto a phosphor screen. The sample was not rotated during these measurements. On the completion of InAs pre-deposition, the As$_2$ source valve was opened at a rate of 40% per second. The integrated intensity of the spot closest to the specular reflection on the QDs' transmission diffraction pattern was monitored by computer and recorded to capture the RHEED signal, $I(t)$. After 60 seconds, the sample was heated to 600 °C for 15 minutes under As$_2$ flux to de-sorb all InAs from the surface, before cooling back down to 475 °C to perform the next RHEED measurement; this procedure allowed all the RHEED measurements to be performed on a single GaAs wafer that could then be re-used.

All the measurements were performed on one day to ensure a stable InAs flux, with a reference signal being collected under "standard" QD growth conditions of continuous InAs deposition at $P_{As} = 6 \times 10^{-6}$ Torr. The peak curvature, $d^2I/dt^2$, of this reference signal was used to measure $\theta_{\text{crit}}$.

A study of triggered QD nucleation as a function of $\theta$ was made by varying the InAs pre-deposition time, including another reference signal based on depositing $\theta < \theta_{\text{crit}}$ to observe the RHEED signal when no QDs were nucleated.

Finally, three repeat measurements under identical growth conditions were performed. No significant variation between these measurements was observed.

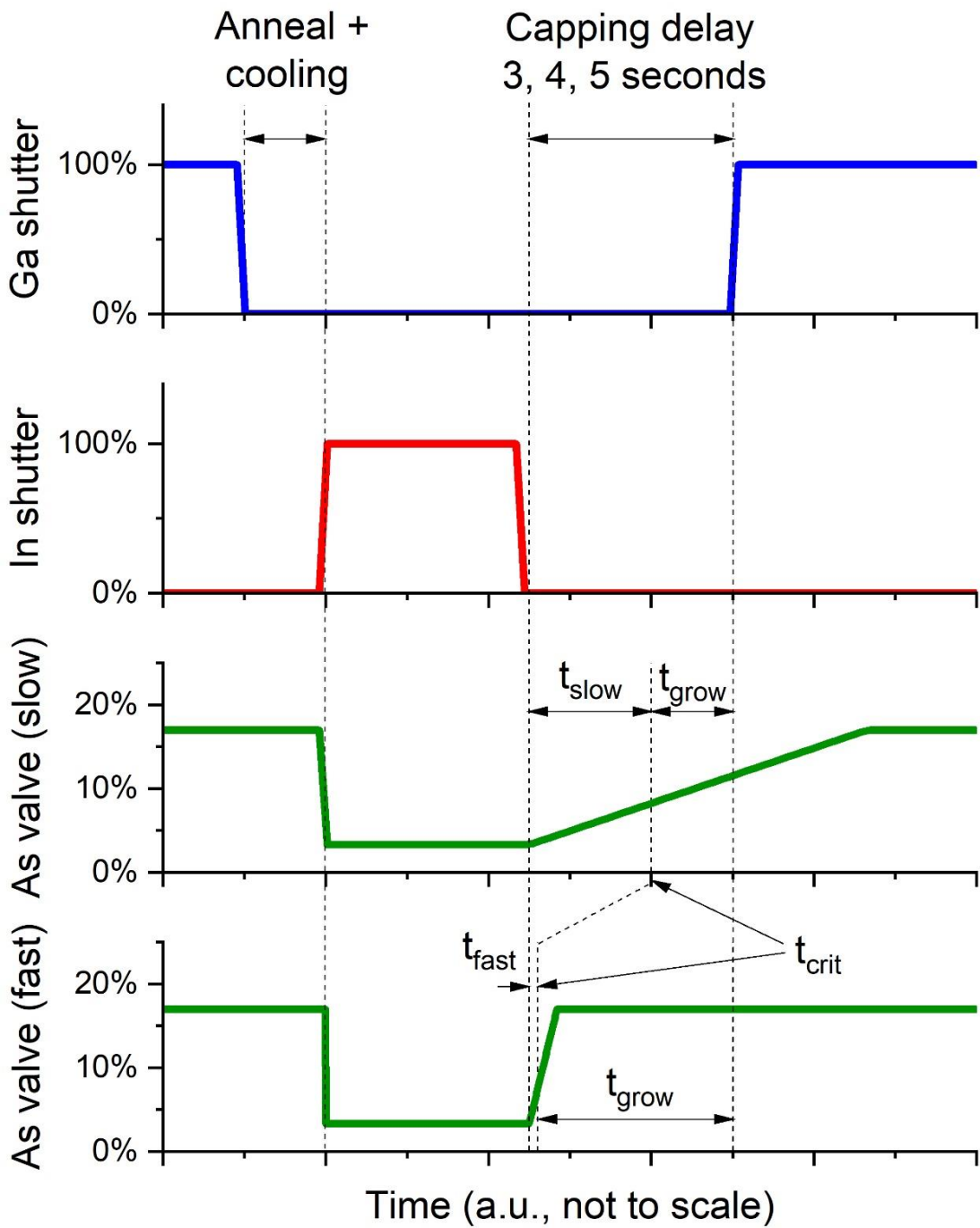

*Figure 6: Top-to-bottom: Ga shutter position (100% = fully open), In shutter position, As valve position for the "slow" sample series, and $As_2$ valve position for the "fast" sample series. After deposition of a GaAs buffer, the surface is smoothed by annealing in an $As_2$ flux prior to the deposition of the InAs layer in an As-poor environment. The As valve is then opened and QD nucleation occurs at $t_{crit}$, labelled as $t_{slow}$ and $t_{fast}$ for the slow and fast series respectively, before being capped by subsequent GaAs deposition. $t_{grow}$ indicates the time between QD nucleation and the start of GaAs overgrowth.*

## PL/AFM sample growth

All stages of sample growth were performed under 30 RPM rotation to ensure uniformity, with no rotation stop for InAs deposition or QD growth and a fixed InAs coverage of $1.2 \times \theta_{\text{crit}}$ was used for all the PL/AFM samples grown for this study. After InAs pre-deposition, the In shutter was closed (Fig. 6b) and the As$_2$ source valve opened; the rate of valve movement was 3.33% per second for the slow series of samples (Fig. 6c) and 40% per second for the fast series (Figure 1d).

The valve movement was monitored via the servo motor's encoder, and a timer started on detection of the valve opening. After the pre-set capping delay time $t$, the Ga shutter was opened by a digital signal (Fig. 6a). This process ensured that highly reliable timing was achieved, avoiding the random delays inherent in the standard serial data communication system employed by the MBE machine. The pre-set delay $t$ was set to 3, 4, or 5 seconds for these samples.

Capping then proceeded at 1 µmhr$^{-1}$ without raising the substrate temperature for the first 10 nm deposition, after which the substrate was ramped to 580 °C for the rest of the GaAs layer. GaAs deposition ceased when the QDs had been buried by a 100 nm thick layer. The surface was then annealed at 580 °C to prepare for a surface layer of uncapped QDs for measurement by AFM, following the identical recipe above, cooling down the sample after the As$_2$ valve is opened.

On completion of growth, the sample was cooled under As$_2$ flux and removed from the MBE reactor.

# Supplementary information
## One-dimensional finite Quantum Well Model

To obtain quantitative data on the early stages of QD growth it was deemed necessary to rely on PL measurements. The main benefit of this approach is a direct measurement of the final properties of the QDs that we wish to exploit, as well as allowing a statistically-significant sample of the total QD population to be obtained in single exposures. For this growth study we were interested in observing the evolution of the QDs' volume as they develop in time.

To estimate the volume of a QD from PL data we neglect in-plane confinement and consider only the confinement in growth direction to obtain $L_z$, which is appropriate given the high aspect ratio of buried InAs/GaAs QDs [16]. This allows an analytical estimation of the lowest energy eigenstate of a one-dimensional QD by modelling it as a one-dimensional finite Quantum Well. It is well-known that this model has no analytical solution for energy $E$ given a well-width $L$, however the inverse problem is fully analytical: this allows us to use a measured $E$ from optical experiments and solve our model to estimate the size of the QD nanocrystal. To begin, we define the confinement energy of electrons $E_{\text{conf}}$ and their binding energy $E_{\text{bind}}$,

$$E_{\text{conf}} = \frac{\Delta E_{\text{E}} (E_{\text{PL}} - E_{\text{InAs}})}{\Delta E_{\text{E}} + \Delta E_{\text{H}}} \quad (3)$$

$$E_{\text{bind}} = \frac{\Delta E_{\text{E}} (E_{\text{WL}} - E_{\text{PL}})}{\Delta E_{\text{E}} + \Delta E_{\text{H}}} \quad (4)$$

where $E_{\text{WL}} = 1.410$ eV and $E_{\text{PL}}$ are the empirically measured energies of the WL and QD emission, respectively. The conduction band offset ratio $\Delta E_{\text{E}}/(\Delta E_{\text{E}} + \Delta E_{\text{H}}) \simeq 5/7$ between electron confinement $\Delta E_{\text{E}}$ and hole confinement $\Delta E_{\text{H}}$ [17] estimates the electron energy because it is more sensitive to QD size than the hole energy because of their lower effective mass. These small QDs are assumed to be pure InAs with bandgap $E_{\text{InAs}} = 0.415$ eV and $m_{\text{e}}^* = 0.023 m_{\text{e}}$ [18, 19]. Solving the Schrödinger equation for $E_{\text{conf}}$ gives,

$$E_{\text{conf}} \simeq \frac{h^2}{8m_e^* L_z^2} \cdot \left( 2/\pi \arctan \sqrt{\frac{E_{\text{bind}}}{E_{\text{conf}}}} \right)^2 \tag{5}$$

Equation (5) is the finite barrier particle-in-a-box ground state solution for one dimension, assuming equal masses in barrier and box. This formula can be re-arranged to give an effective volume $V_{\text{eff}}$ of the QD according to the electron's confinement, by introducing a parameter $\alpha$ to account for the aspect ratio of the real nanostructure:

$$V_{\text{eff}} = \alpha L_z^3 \simeq \alpha \left( \sqrt{\frac{2\hbar^2}{m_e^* E_{\text{conf}}}} \tan^{-1} \sqrt{\frac{E_{\text{bind}}}{E_{\text{conf}}}} \right)^3 \tag{6}$$

The accuracy of Eq. (6) is dependent on the choice of $m_e^*$ and $E_{\text{InAs}}$, and the result depends on the aspect ratio of the quantum dot, which is described by $\alpha$. As we are interested in the shape of the volume spectrum, $m_e^*$ is not important, while the chosen value of $E_{\text{InAs}}$ changes $V_{\text{eff}}$ in proportion, to a first order expansion.

Additional effects not accounted for in this model are that of the binding energy due to Coulomb or exchange, or other correlation effects; these effects can be very strong in these QDs, with shifts on the order of 1-20 meV [20], however they will be a smooth function of QD size and therefore are expected to cause limited distortion to the results. Moreover, the primary purpose of this model is to assess and design sample growth procedures: it benefits both from the simplicity of the model and its empirical nature, while at the same time not claiming to give a precise and absolute measure of QD dimensions.

### Gamma distribution and nanoparticle growth

The ensemble properties of InAs/GaAs QDs are often interpreted using the Gaussian distribution to extract the intensity, centroid wavelength and full-width at half-maximum. Gaussian distributions often arise in situations where an equilibrium is achieved with both positive and negative random fluctuations summing together, creating a distribution around a mean balance. In the early stages of growth however there is no such equilibrium and a Gaussian distribution is not appropriate because the dominant contributions are from only positive fluctuations: atoms being incorporated into the growing QDs at random under diffusion. Effects that limit the development of the QDs such as evaporation, self-size limiting due to strain, or exhaustion of available adatoms are not significant at first. Under these circumstances a more appropriate distribution is the Gamma distribution,

$$f(x, \alpha, \beta) = \frac{\beta^\alpha}{\Gamma(\alpha)} x^{\alpha-1} e^{-\beta x} \tag{7}$$

which arises when $x$ is the summation of several exponentially, or half-normally distributed positive contributions. For growth of nanocrystals on a two-dimensional surface where diffusion is controlled by the distances to the nearest neighbouring (competing) nanostructures this is the appropriate case, as shown by Kiang [14], then given analytical foundation by Weaire *et al.* [15]. These studies determined that for the case of completely random and uncorrelated particle positions, a gamma distribution with $\beta = 3.61$ is appropriate. For situations where the particles have some correlations, for example due to their finite size creating a minimum nearest-neighbour distance, $\beta$ increases [21].

Measurements of InAs/GaAs QDs by AFM/STM microscopy have determined that some correlations appear to exist: Gamma distributions with $\beta > 3.61$ were determined by Fanfoni *et al.* [22] for their distributions of QD volumes and their associated Voronoi cell areas, although they do not show a direct correlation of these two datasets. Our own measurements confirm their presented data but find no correlation between *final* QD volume and the associated Voronoi cell area, even though they have identical distributions. The Hopkins-Skellam index has also been used to infer the presence of correlation in the positions of QDs [10]. Correlations that increase $\beta$ would be disadvantageous to the growth procedure outlined in this paper because the long-tail of large QDs would be increasingly suppressed, making it harder to resolve the optical transitions of individual QDs.

A second point of interest in the gamma distribution is whether the QD growth in its earliest stages is due to growth by diffusion causing competition with the nanostructures nearest neighbours. If this were the case then the largest QDs would be distant from any neighbouring nanostructures, reducing the effect of quantum tunnelling and Coulomb charge noise between QDs. Therefore, the large QDs in the tail of the gamma distribution would likely exhibit very good coherence properties, even in a sample with a high total density of QDs.

## Fitting $t_{\text{crit}}$ and growth rate

The As$_2$ BEP is controlled by a valve in the molecular beam source, where the BEP $P_{\text{As}}$ and the valve position $y$ are related by a calibration curve $P_{\text{As}} = \aleph(y)$, shown in Figure 3(d), that is derived from measurements at the start of a growth day. During InAs deposition, the valve was restricted to a position $y_0 = 3.3\%$ for $P_{\text{As}} = 1.1 \times 10^{-7}$ Torr, then we define $t = 0$ by the detection of valve movement as it is opened to the nominal BEP for GaAs growth, $P_{\text{As}} = 6.0 \times 10^{-6}$ Torr, after which $P_{\text{As}}$ evolved as:

$$P_{As}(t) = \aleph(R_{\text{valve}}t + y_0), \tag{8}$$

As we are controlling the surface energy $S$ by changing $P_{\text{As}}$ we can define $P_{\text{crit}}$ as the pressure when QD nucleation is triggered. With knowledge of $P_{\text{crit}}$ we can calculate the timing of QD nucleation $t_{\text{crit}}$ and vice-versa:

$$t_{\text{crit}} = \frac{\aleph^{-1}(P_{\text{crit}}) - y_0}{R_{\text{valve}}} \tag{9}$$

To determine these values, we set the criterion that both the fast and slow mean QD volume vs. $t$ data can be fitted with a straight-line of the same gradient but different intercepts $t_{\text{fast}}$ and $t_{\text{slow}}$ respectively (Figure 3(c)); least squares fitting is performed simultaneously to determine the best mutual fit with the constraint that the intercepts are related by:

$$t_{\text{fast}} = \frac{R_{\text{slow}}}{R_{\text{fast}}} t_{\text{slow}} \tag{10}$$

Where $R_{\text{slow}}/R_{\text{fast}}$ is the ratio of valve movement rates for the fast and slow samples. Fitting to the mean QD volume yields, $t_{\text{slow}} = (1.59 \pm 0.07)$ s implying a critical arsenic flux $P_{\text{crit}} = 1.34 \times 10^{-6}$ Torr, and a growth rate of $(0.179 \pm 0.004)$ nm$^3$s$^{-1}$ for the QDs of mean size, assuming they are pure InAs and the aspect ratio $\alpha = 1$. The actual growth rate is significantly larger due to the shape of a real QD having $\alpha \gg 1$ and In-Ga intermixing inside the QD. The uncertainty over the absolute growth rate of the nanocrystal itself is therefore substantial but the relative uncertainty between samples is small.

The clear linear trendline intercepting at $t = 0$ is consistent with growth by pure kinetic diffusion, demonstrating that QD growth is deterministic and can be controlled precisely if the initial conditions are also controlled; this offers the possibility of combining site-control by substrate patterning with this triggering technique to create arrays of identical QDs.

## Predictive design model for QD growth

With the knowledge of the statistical population of QD volumes derived from wavelength (dashed-line of Figure 7(a)), and the mean growth rate of volume with time, it is now possible to predict the growth time required to achieve a given wavelength and density of QDs at that wavelength.

Using the survival function for the gamma distribution with $\beta = 3.61$ (solid line of Figure 7(a)) and the total QD density in the layer from AFM measurements, the target density can be expressed in terms of the target volume of QD, in units of mean QD volume. This target density is defined here as the combined density of all QDs equal to or larger than the volume indicated; this definition is useful because QDs at longer wavelengths may cause parasitic absorption or emission at the target wavelength. QDs at shorter wavelengths are less problematic; it is important to recognise that this growth procedure yields a consistent total density of QDs of all volumes and that we are calculating the spectral density of the larger QDs whose emission can be spectrally-isolated from the others.

The volume of a QD emitting at the desired wavelength can be calculated from the model above, allowing the target mean volume to be deduced.

Dividing the target mean volume by the mean growth rate of the QDs, then adding $t_{\text{crit}}$, gives the required time delay between opening the As$_2$ source valve and initiating GaAs overgrowth to bury the QD layer. The results of using this procedure are plotted in Figure 7(b), with the curves chosen for different target densities allowing the growth time to be read-off for a desired wavelength; the growth times are determined for opening the valve at the slow rate.

## Photoluminescence model for exciton – biexciton power dependence

The power dependence of the discrete PL emission lines can be used to identify whether they are exciton or biexciton states: the former will have a linear dependence on excitation power at low excitation, while the latter will have a parabolic dependence [23]. At higher excitation power the exciton and biexciton emission lines will saturate and then decrease in intensity as more electron-hole pairs occupy the QD confining potential. To capture this characteristic behaviour, we used a simple rate equation model.

In the model, an excitation $I$ is supplied to the wetting layer, represented by the number density of carriers $N_{\text{WL}}$, from where it is destroyed either by radiative recombination with lifetime $\tau_{\text{WL}}$, or capture into a QD with lifetime $\tau_c$. $N_{\text{WL}}$ is treated as an infinite reservoir with no upper limit to its population. The average QD microstate is described by $n$, where $n = 0$ is an empty dot, $n = 1$ is a dot with a single exciton, $n = 2$ is a biexciton, and $n = 3$ denotes QDs in all higher excited states. $N_n$ denote the probability of finding a QD in a given microstate, with an upper-limit of 1 defined by the initial conditions, and $\tau_{rn}$ being the radiative lifetime of each microstate. The model is similar to other rate equation modelling, such as a random population model, or multi-exciton complexes [24, 25, 23] except that it does not consider charged QDs, only the neutral excitonic states, in order to capture the basic exciton-biexciton behaviour with the minimum of input parameters. Therefore, we define a system of differential equations:

$$\frac{dN_{\text{WL}}}{dt} = I - \frac{{N_{WL}}^2}{\tau_{\text{WL}}} - \sum_{n=0,1,2} \frac{N_{\text{WL}} N_n}{\tau_c}$$

$$\frac{dN_0}{dt} = \frac{N_1}{\tau_{r1}} - \frac{N_{\text{WL}} N_0}{\tau_c}$$

$$\frac{dN_1}{dt} = \frac{N_2}{\tau_{r2}} + \frac{N_{\text{WL}} N_0}{\tau_c} - \frac{N_{\text{WL}} N_1}{\tau_c} - \frac{N_1}{\tau_{r1}}$$

$$\frac{dN_2}{dt} = \frac{N_3}{\tau_{r3}} + \frac{N_{WL}N_1}{\tau_c} - \frac{N_{WL}N_2}{\tau_c} - \frac{N_2}{\tau_{r2}}$$

$$\frac{dN_3}{dt} = \frac{N_{WL}N_2}{\tau_c} - \frac{N_3}{\tau_{r3}}$$

The initial conditions are that all QDs are empty, $N_0 = 1$ and $N_{n=1,2,3} = 0$, and that the wetting layer is de-populated: $N_{WL} = 0$. For each excitation $I$, the system of equations is solved for the steady-state solution with respect to time; the outputs are then the radiative recombination rates of the exciton and biexciton states $N_1/\tau_{r1}$ and $N_2/\tau_{r2}$, respectively.

The biexciton recombination lifetime $\tau_{r2} = \tau_{r1}/2$ to reflect the double likelihood of recombination with two electron-hole pairs present. For simplicity the radiative lifetime of "all other microstates" $\tau_{r3} = \tau_{r2}$ and this assumption allows the model to capture the qualitative quenching behaviour. The fit presented in the main paper sets the other parameters as: $\tau_{WL} = 5$, $\tau_{r1} = 50$, $\tau_c = 11$, and $I$ was calculated from 1×10$^{-5}$ to 0.15 all in arbitrary units and the results were linearly scaled to fit the experimental data.

In conclusion, this simplified model was able to capture the power-law excitation dependence, as well as the characteristic saturation and onset of quenching; the model also has qualitative agreement with the full quenching behaviour. Accurate modelling of the exact carrier dynamics requires a more detailed model, at the expense of requiring a greater amount of knowledge about the QD in question. Therefore, it is useful as a quick way of verifying exciton-biexciton behaviour from individual InAs/GaAs QDs.

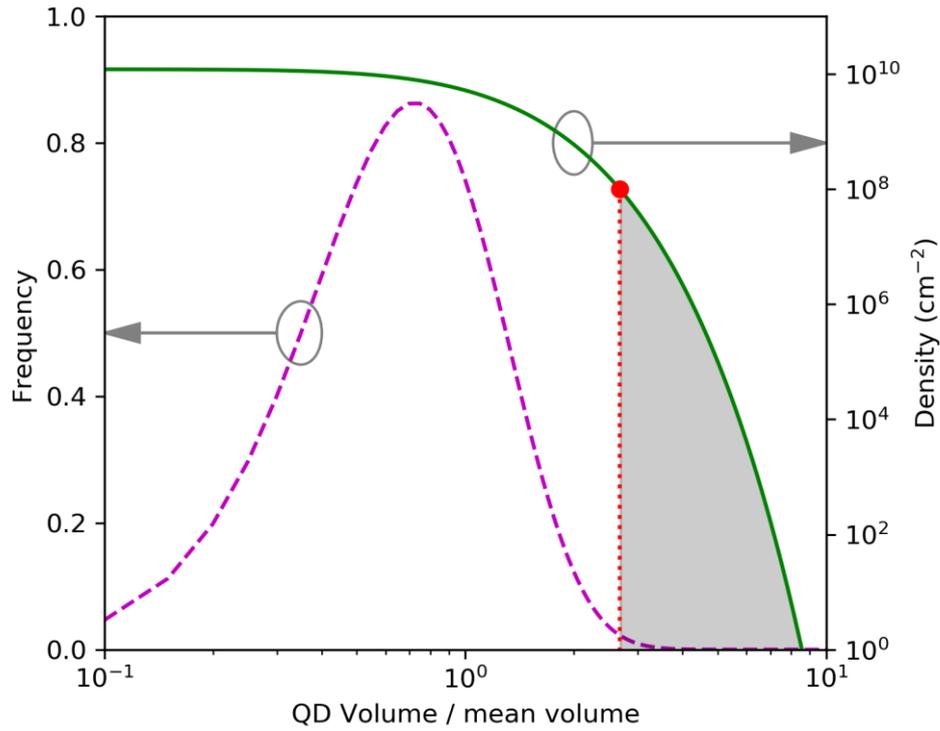

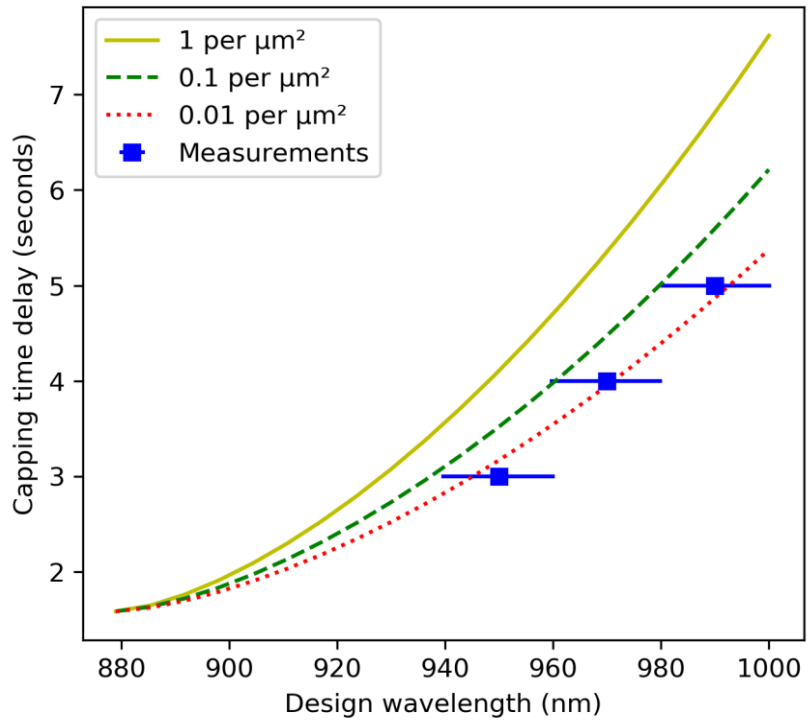

*Figure 7: (a) Dashed line is the gamma distribution for $\beta = 3.61$, while the solid green line is the survival function for this distribution scaled by the total density of QDs measured by AFM. The red line indicates how the target density is used to select the desired QD volume. (b) The results of applying this process to a range of wavelengths, for three desired target densities. The blue points indicate measurements taken on several samples (see main paper).*

# References

[1] D. Ritchie, R. Stevenson, R. Young, A. Hudson, C. Salter, D. Ellis, A. Bennett, P. Atkinson, K. Cooper, I. Farrer, C. Nicoll and A. Shields, "Single-photon and entangled-photon sources for quantum information," in *2010 Conference on Optoelectronic and Microelectronic Materials and Devices*, Canberra, ACT, Australia, 2010.

[2] P. Senellart, G. Solomon and A. White, "High-performance semiconductor quantum-dot single-photon sources," *Nature Nanotechnology,* vol. 12, pp. 1026-1039, 2017.

[3] C. L. Salter, R. M. Stevenson, I. Farrer, C. A. Nicoll, D. A. Ritchie and A. J. Shields, "An entangled-light-emitting diode," *Nature,* vol. 465, p. 594, 2010.

[4] R. Dingle, W. Weigmann and C. H. Henry, "Quantum States of Confined Carriers in Very Thin AlxGa1-xAs-GaAs- AlxGa1-xAs Heterostructures," *Physical Review Letters,* vol. 33, p. 827, 1974.

[5] E. Bauer, "Phänomenologische Theorie der Kristallabscheidung an Oberflächen. I," *Zeitschrift für Kristallographie,* vol. 110, pp. 372-394, 1958.

[6] F. Leroya, Ł. Borowikbc, F. Cheynisa, Y. Almadoribc, S. Curiottoa, M. Trautmanna, J. Barbébc and P. Müller, "How to control solid state dewetting: A short review," *Surface Science Reports,* vol. 71, no. 2, pp. 391-409, 2016.

[7] I. N. Stranski and L. Krastanow, "Zur Theorie der orientierten Ausscheidung von Ionenkristallen aufeinander," *Monatshefte für Chemie,* vol. 71, p. 351, 1937.

[8] K. H. Schmidt, G. Medeiros-Ribeiro, J. Garcia and P. M. Petroff, "Size quantization effects in InAs self-assembled quantum dots," *Applied Physics Letters,* vol. 70, no. 13, pp. 1727-1729, 1997.

[9] P. Kratzer, E. Penev and M. Scheffler, "Understanding the growth mechanisms of GaAs and InGaAs thin films by employing first-principles calculations," *Applied Surface Science,* vol. 216, pp. 436-446, 2003.

[10] T. Konishi, G. R. Bell and S. Tsukamoto, "Hopkins-Skellam index and origin of spatial regularity in InAs quantum dot formation on GaAs(001)," *Journal of Applied Physics,* vol. 117, p. 144305, 2015.

[11] R. P. Mirin, A. Roshko, M. v. d. Puijl and A. G. Norman, "Formation of InAs/GaAs quantum dots by dewetting during cooling," *Journal of Vacuum Science and Technology B,* vol. 20, p. 1489, 2002.

[12] O. Gywat, H. J. Krenner and J. Berezovsky, Spins in optically active quantum dots: concepts and methods, John Wiley & Sons., 2009.

[13] A. J. Bennett and R. Murray, "Nucleation and ripening of seeded InAs/GaAs quantum dots," *Journal of Crystal Growth,* vol. 240, pp. 436-444, 2002.

[14] T. Kiang, "Random Fragmentation in Two and Three Dimensions," *Zeitschrift für Astrophysik,* vol. 64, pp. 433-439, 1966.

[15] D. Weaire, J. P. Kermode and J. Wejchert, "On the distribution of cell areas in a Voronoi network," *Philosophical Magazine B,* vol. 53, pp. L101-L105, 1986.

[16] O. Gywat, H. J. Krenner and J. Berezovsky, Spins in optically active quantum dots: concepts and methods, John Wiley & Sons, 2009.

[17] E. Harbord, P. Spencer, E. Clarke and R. Murray, "Radiative lifetimes in undoped and p-doped InAs/GaAs quantum dots," *Physical Review B,* vol. 80, p. 195312, 2009.

[18] Z. Fang, K. Ma, D. Jaw, R. Cohen and G. Stringfellow, "Photoluminescence of InSb, InAs, and InAsSb grown by organometallic vapor phase epitaxy," *Journal of Applied Physics,* vol. 67, no. 11, pp. 7034-7039, 1990.

[19] F. P. Kesamanly, Y. V. Mal'tsev, D. N. Nasledov, L. A. Nikolaeva, M. N. Pivovarov, V. A. Skripkin and Y. I. Ukhanov, "Structure of the conduction band in indium arsenide," *Fizika i Tekhnika Poluprovodnikov,* vol. 3, no. 8, p. 1182, 1969.

[20] M. Grundmann, O. Stier and D. Bimberg, "InAs/GaAs pyramidal quantum dots: Strain distribution, optical phonons, and electronic structure," *Physical Review B,* vol. 52, no. 16, pp. 11969-11981, 1995.

[21] P. A. Mulheran and J. A. Blackman, "The origins of island size scaling in heterogeneous film growth," *Philosophical Magazine Letters,* vol. 72, no. 1, pp. 55-60, 1995.

[22] M. Fanfoni, E. Placidi, F. Arciprete, E. Orsini, F. Patella and A. Balzarotti, "Sudden nucleation versus scale invariance of InAs quantum dots on GaAs," *Physical Review B,* vol. 75, p. 245312, 2007.

[23] J. J. Finley, A. D. Ashmore, A. Lemaître, D. J. Mowbray, M. S. Skolnick, I. E. Itskevich, P. A. Maksym, M. Hopkinson and T. F. Krauss, "Charged and neutral exciton complexes in individual self-assembled In(Ga)As quantum dots," *Physical Review B,* vol. 63, p. 073307, 2001.

[24] M. Grundmann and D. Bimberg, "Theory of random population for quantum dots," *Physical Review B,* vol. 55, no. 15, pp. 9740-9745, 1997.

[25] E. Dekel, D. Gershoni, E. Ehrenfreund, J. M. Garcia and P. M. Petroff, "Carrier-carrier correlations in an optically excited single semiconductor quantum dot," *Physical Review B,* vol. 61, p. 11009, 2000.